\documentclass[prl,twocolumn,showpacs]{revtex4}
\usepackage{graphicx}
\begin{document}
\title
{Localized growth modes, dynamic textures, and upper critical
dimension for the Kardar-Parisi-Zhang equation in the weak noise
limit}
\author{Hans C. Fogedby}
\email{fogedby@phys.au.dk} \affiliation {Department of Physics and
Astronomy,
University of Aarhus DK-8000, Aarhus C, Denmark\\
and
\\
NORDITA, Blegdamsvej 17,
DK-2100, Copenhagen {\O}, Denmark }
%\date{\today}
\begin{abstract}
A nonperturbative weak noise scheme is applied to the
Kardar-Parisi-Zhang equation for a growing interface in all
dimensions. It is shown that the growth morphology can be
interpreted in terms of a dynamically evolving texture of
localized growth modes with superimposed diffusive modes. Applying
Derrick's theorem it is conjectured that the upper critical
dimension is four.
\end{abstract} \pacs{05.10.Gg, 05.45.-a, 64.60.Ht, 05.45.Yv}
\maketitle

There is a current interest in the general morphology and in
particular the scaling properties of nonequilibrium models in
statistical physics. There is, moreover, a need to develop general
methods beyond perturbation theory, renormalization group theory,
mode coupling theory, and numerical simulations, which permit an
analysis of these often intractable problems.

The purpose of this Letter is two-fold. On the one hand, I should
like to draw the attention of the community to the availability of
a weak noise approach to stochastic equations driven by Gaussian
noise which is based on a principle of least action and which
allows a discussion of stochastic processes in terms of classical
equations of motion and, moreover, provides the Arrhenius factor
or weight of a specific kinetic transition; the method is
summarized below. Secondly, I apply the weak noise approach to the
Kardar-Parisi-Zhang (KPZ) equation for the kinetic growth of an
interface in arbitrary dimensions and show that the growth
morphology can be interpreted as a dynamical network of growth
modes with superimposed diffusive modes. As a corollary, using
Derrick's theorem, it is finally shown that the texture of growth
modes does not persist above four dimensions, indicating that the
upper critical dimension of the KPZ equation is $d=4$. Details and
further developments will be discussed elsewhere.

The variationally-based weak noise method dates back to Onsager
\cite{Onsager53} and has since reappeared as the Freidlin-Wentzel
theory of large deviations \cite{Freidlin98,E04,Graham89} and as
the weak noise saddle point approximation to the functional
Martin-Siggia-Rose scheme \cite{Martin73,Baussch76}. The point of
departure is the Langevin equations for the set of stochastic
variables $w_n$, $n=1,\cdots N$, driven by white Gaussian noise
\cite{vanKampen92,Stratonovich63}
\begin{eqnarray}
&&\frac{dw_p}{dt}=-\frac{1}{2}F_p-\frac{\Delta}{2}G_{mn}\nabla_mG_{pn}
+G_{pn}\eta_n, \label{lan}
\\
&&\langle\eta_n\eta_m\rangle(t)=\Delta\delta_{nm}\delta(t),
\label{noise}
\end{eqnarray}
where $F_n(w_p)$ is the drift, $G_{nm}(w_p)$ is accounting for
multiplicative noise, $\nabla_n=\partial/\partial w_n$ and
$\Delta$ is the explicit noise strength; sums are performed over
repeated indices. The associated Fokker-Planck equation for the
probability distribution $P(w_n,t)$ then has the form
\cite{vanKampen92,Stratonovich63}
\begin{eqnarray}
\frac{\partial P}{\partial
t}=\frac{1}{2}\nabla_n[F_n+\Delta\nabla_mK_{mn}]P, \label{fp}
\end{eqnarray}
where the symmetrical noise matrix
$K_{pm}(w_p)=G_{pn}(w_p)G_{mn}(w_p)$.

Introducing the WKB ansatz
\begin{eqnarray}
P(w_n,t)\propto\exp[-S(w_n,t)/\Delta], \label{wkb}
\end{eqnarray}
the Hamilton-Jacobi equation $\partial S/\partial t+H=0$,
$p_n=\nabla_nS$ follows to leading order in $\Delta$ with
Hamiltonian
\begin{eqnarray}
H=-\frac{1}{2}p_nF_n+\frac{1}{2}K_{nm}p_np_m, \label{ham}
\end{eqnarray}
The Hamilton equations of motion are
\begin{eqnarray}
&&\frac{dw_n}{dt}=-\frac{1}{2}F_n + K_{nm}p_m, \label{eq1}
\\
&&\frac{dp_n}{dt}=\frac{1}{2}p_m\nabla_nF_m
-\frac{1}{2}p_mp_q\nabla_nK_{mq}, \label{eq2}
\end{eqnarray}
determining classical orbits on the energy surfaces given by $H$
in a classical $(w_n,p_n)$ phase space. Finally, the action $S$ is
given by
\begin{eqnarray}
S(w_n,T)=\int^{w_n,T}dtd^Nw~p_n\frac{dw_n}{dt} - HT. \label{act}
\end{eqnarray}
The weak noise scheme bears the same relationship to stochastic
fluctuations as the WKB approximation in quantum mechanics,
associating the phase of a wave function with the action of a
classical orbit. In addition to providing a classical orbit
picture of stochastic fluctuations and thus allowing the use of
dynamical system theory, the method also yields the Arrhenius
factor $P\propto\exp(-S/\Delta)$ for a kinetic transition to $w_n$
in time $T$. Here the action $S$ serves as the weight in the same
manner as the energy $E$ in the Boltzmann factor
$P\propto\exp(-E/kT)$ for equilibrium processes. This completes
the brief review of the nonperturbative weak noise scheme; for
details see e.g. Ref. \cite{Fogedby99a}

The KPZ equation is a field theoretic Langevin equation describing
the nonequilibrium growth of an interface
\cite{Kardar86,Medina89,Halpin95,Krug97}
\begin{eqnarray}
&&\frac{\partial h}{\partial t}=\nu\nabla^2h
+\frac{\lambda}{2}\nabla h\cdot\nabla h-F+\eta, \label{kpz}
\\
&&\langle\eta\eta\rangle ({\bf r},t)=\Delta\delta^d({\bf
r})\delta(t). \label{noise2}
\end{eqnarray}
Here $h({\bf r},t)$ is the height of the growth profile, $\nu$ a
diffusion coefficient, $\lambda$ a growth coefficient, $F$ an
imposed drift, and $\eta$ a locally correlated white Gaussian
noise of strength $\Delta$. Dynamic renormalization group (DRG)
studies \cite{Forster76,Forster77,Medina89} indicates that the KPZ
equation conforms to the dynamical scaling hypothesis with long
time -long distance correlations $\langle h h\rangle({\bf r},t)
=r^{2\zeta}\Phi(t/r^z)$, characterized by roughness exponent
$\zeta$, dynamic exponent $z$ and scaling function $\Phi$. The KPZ
equation is, moreover, invariant under the Galilean
transformation, ${\bf r}\rightarrow {\bf r}-\lambda {\bf u^0}t$,
$h\rightarrow h+{\bf u^0}\cdot{\bf r}$, $F\rightarrow
F+(\lambda/2){\bf u^0}\cdot{\bf u}^0$, and the slope $\nabla
h={\bf u}\rightarrow{\bf u}+{\bf u}^0$ which, implying the scaling
law $\zeta+z=2$, delimits the KPZ universality class. In $d=1$ the
stationary distribution of $h$ is
$P(h)\propto\exp[-(\nu/\Delta)\int dx~(\nabla h)^2]$
\cite{Huse85}, yielding $\zeta=1/2$ and $z=2-\eta=3/2$. In $d\ge
2$ the DRG implies a kinetic phase or roughness transition at a
finite $\lambda$ from a smooth phase with $z=2$, and
$\zeta=(2-d)/2$, the linear Edwards-Wilkinson (EW) case for
$\lambda=0$ \cite{Edwards82}, to a rough phase with still
debatable scaling exponents $z$ and $\zeta=2-z$, see e.g. Refs.
\cite{Kim89,Wolf97,Laessig98a,Fogedby02c}. It has, moreover, been
conjectured that $d=4$ is an upper critical dimension beyond which
the KPZ equation exhibits EW behavior
\cite{Laessig97,Moore95,Colaiori01a}.

Applying the nonlinear Cole-Hopf transformation \cite{Medina89}
$w({\bf r},t)= \exp(\lambda h({\bf r},t)/2\nu)$ the KPZ equation
(\ref{kpz}) takes the form
\begin{eqnarray}
\frac{\partial w}{\partial t} = \nu\nabla^2w
-\frac{\lambda}{2\nu}wF+\frac{\lambda}{2\nu}w\eta, \label{ch}
\end{eqnarray}
with multiplicative noise. In a moving frame $w$ is changed
according to $w\rightarrow w\exp[(\lambda/2\nu){\bf u}^0\cdot{\bf
r}]$. Note that in the noiseless case for $\eta=0$ (\ref{ch})
reduces to the linear diffusion equation permitting a rather
complete analysis of the deterministic KPZ equation for a relaxing
interface \cite{Woyczynski98,Medina89}. The Cole-Hopf equation
(\ref{ch}) with multiplicative noise forms the basis for the
mapping of the KPZ equation to a model of directed polymers (DP)
in a quenched random medium which by means of the replica method
relates the roughness transition to a pinning transition in the DP
model \cite{Kardar87b,Kardar87a}.

In recent work I have applied the weak noise method to the KPZ
equation or, equivalently, the noisy Burgers equation in $d=1$
\cite{Fogedby99a,Fogedby03b}. Here the method readily yields a
consistent Galilean invariant dynamical picture of a growing
interface in terms of propagating domain walls with superimposed
diffusive modes. The localized domain walls account for the
growth; the diffusive modes form a subdominant background governed
by EW dynamics and not contributing to the growth. The method,
moreover, identifies scaling exponents, universality classes, and
associates the dynamic exponent $z$ with the domain wall
dispersion law or the Hurst exponent $H=1/z$ with the anomalous
superdiffusion of growth modes, see also Ref. \cite{Halpin95}.

Drawing on the insight gained in $d=1$, the weak noise method is
here applied to the KPZ equation in arbitrary dimensions
\cite{Fogedby02c}. Unlike the $d=1$ case, where the local slope
field $u=\nabla h$ is the natural variable, the Cole-Hopf
diffusive field $w$ is more convenient for analysis in $d>1$.
Referring to the general weak noise scheme and making the
assignment, $w_n\rightarrow w$, $p_n\rightarrow p$,
$K_{nm}\rightarrow(\lambda/2\nu)^2w^2\delta^d({\bf r}-{\bf r}')$,
and $F_n\rightarrow -2[\nu\nabla^2w-(\lambda/2\nu)wF]$, the
Hamiltonian density is given by
\begin{eqnarray}
{\cal H}=p[\nu\nabla_n\nabla_n-\nu k^2]w+\frac{1}{2}k_0^2w^2p^2,
\label{ham2}
\end{eqnarray}
and the canonical field equations take the form
\begin{eqnarray}
&&\frac{\partial w}{\partial t} = \nu\nabla^2w -\nu
k^2w+k_0^2w^2p, \label{eq11}
\\
&&\frac{\partial p}{\partial t} =- \nu\nabla^2p +\nu
k^2p-k_0^2p^2w. \label{eq22}
\end{eqnarray}
Here $k=(\lambda F/2\nu^2)^{1/2}$ and $k_0=\lambda/2\nu$ define
two characteristic inverse length scales. Using the equation of
motion (\ref{eq11}) the action is
\begin{eqnarray}
S(w,T)=\frac{1}{2}k_0^2\int_{w_i,0}^{w,T} d^dxdt~w^2p^2,
\label{act2}
\end{eqnarray}
and $P(w,T)\propto\exp(-S(w,T)/\Delta)$ yields the transition
probability.

The action and as a result the equations of motion are invariant
under the combined Galilei transformation $w\rightarrow w
\exp[(\lambda/2\nu){\bf u}^0\cdot{\bf r}]$, $p\rightarrow
p\exp[-(\lambda/2\nu){\bf u}^0\cdot{\bf r}]$. The equations of
motion determine orbits in the canonical $(w,p)$ phase space from
an initial configuration $w_i$ to a final configuration $w$
traversed in time $T$ with the noise field $p$ as a slaved
variable. The orbits lie on the constant energy surfaces $H=\int
d^dx~{\cal H}$. The action evaluated along the orbit then yields
the transition probability $P(w_i\rightarrow w,T)$. The formalism
is symplectic (canonical) and it is an easy task to perform
canonical transformations to the height field $h$ or the local
slope field ${\bf u}=\nabla h$.

The growth of the interface is due to the propagation of localized
modes across the system. As in the $d=1$ case the first task is
thus to identify the relevant excitations and connect them in a
dynamical network in order to obtain a consistent growth
morphology. In the static limit the field equations (\ref{eq11})
and (\ref{eq22}) assume the symmetrical form: $\nu\nabla^2w=\nu
k^2 w-k_0^2w^2p$ and $\nu\nabla^2p=\nu k^2 p-k_0^2p^2w$. On the
noiseless manifold, $p=0$, and the  noisy manifold, $p=\nu w$, the
static equations reduce to the linear diffusion equation and the
nonlinear Schr\"odinger equation (NLSE), respectively,
\begin{eqnarray}
&&\nabla^2 w = k^2 w~~~~~~~~~~~~\text{for}~~p=0~, \label{diff}
\\
&&\nabla^2 w = k^2 w -k_0^2 w^3~~\text{for}~~p=\nu w. \label{nlse}
\end{eqnarray}

The diffusion equation (\ref{diff}) admits the radially symmetric
solution $w_+(r)\propto r^{1-d/2}I_{1-d/2}(kr)$, where $I(z)$ is
the modified Bessel function \cite{Lebedev72}. For small $r$,
$w_+(r)\rightarrow \text{cst}$. For large $r$, $w_+(r)\propto
r^{(1-d)/2}\exp(kr)$, yielding the asymptotic height field
$h_+(r)=(2\nu/\lambda)((1-d)/2)\ln r+kr)\approx (2\nu/\lambda)kr$,
and the outward-pointing vector slope field ${\bf
u}_+(r)=(2\nu/\lambda)k{\bf r}/r$ of constant magnitude
$(2\nu/\lambda)k$. In $d=1$, $w_+(x)\propto\cosh(kx)$, giving rise
to the height field $h_+(x)=(2\nu/\lambda)\ln\cosh(kx)$, and the
slope field $u_+(x)=\nabla h=(2\nu/\lambda)k\tanh(kx)$, i.e., the
right hand domain wall solution of the static Burgers equation
$\nu\nabla^2u=-\lambda u\nabla u$ \cite{Burgers74,Fogedby98a}.
Since the solution  $w_+$ lives on the noiseless manifold $p=0$
and the action $S_+=0$ it carries no dynamics.

\begin{figure}
\includegraphics[width=.9\hsize]
{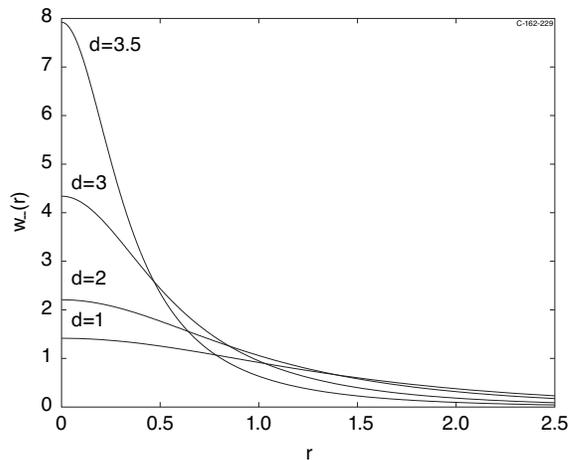}
%{c:/user/manus/figs/162-229.eps}
\caption{The
radially symmetric bound states of the NLSE are depicted for
$k=k_0=1$ in $d=1, d=2, d=3, d=3.5$.} \label{fig1}
\end{figure}

The NLSE (\ref{nlse}) admits a radially symmetric bound state
$w_-(r)$ falling off as $w_-(r)\propto r^{(1-d)/2}\exp(-kr)$ for
large $r$ \cite{Finkelstein51,Chiao64,Rasmussen86}. At the origin
$w_-(0)=a_dk/k_0$, where $a_d$ depends on the dimension.
Numerically, $a_1=1.41 (\sqrt 2)$, $a_2=2.21$, $a_3=4.34$. In the
limit $d\rightarrow 4$ the amplitude diverges, the width vanishes,
and the bound state disappears for $d\ge 4$. The asymptotic height
field $h_-(r) = -(2\nu/\lambda)kr$ and the inward-pointing slope
field ${\bf u}_-=-(2\nu/\lambda)k{\bf r}/r$ of constant magnitude
$(2\nu/\lambda)k$. In $d=1$ the NLSE admits the soliton solution
$w_-(x)=a_1(k/k_0)\cosh^{-1}(kr)$ yielding the height field
$h_-(r)=-(2\nu/\lambda)\ln\cosh(kr)$ and slope field $u_-=\nabla
h_-=-(2\nu/\lambda)k\tanh(kr)$, i.e., the noise-induced left hand
domain wall solution \cite{Fogedby95,Fogedby98c}. Here the
solution $w_-$ is associated with the noisy manifold $p=\nu w$ and
carries according to (\ref{act2}) the action $S_-=(k_0^2/2)T\int
d^dx~w_-^4$. In Fig.~\ref{fig1} the radial solutions of the NLSE
are depicted with parameter choice $k=1$ and $k_0=1$ in $d=1, 2,
3, 3.5$.

At large distances the slope fields ${\bf u}_\pm$ associated with
$w_\pm$ approach vector fields of constant magnitude
$2(\nu/\lambda)k$ and the boundary condition ${\bf u}=0$,
corresponding to a flat interface, is implemented by combining a
set of modes with appropriately chosen amplitudes $k$. In a charge
language, connecting modes with positive ($k>0$) and negative
charges ($k<0$) and enforcing charge neutrality $\sum_i k_i=0$,
the boundary condition is automatically enforced.

\begin{figure}
\includegraphics[width=.9\hsize]
{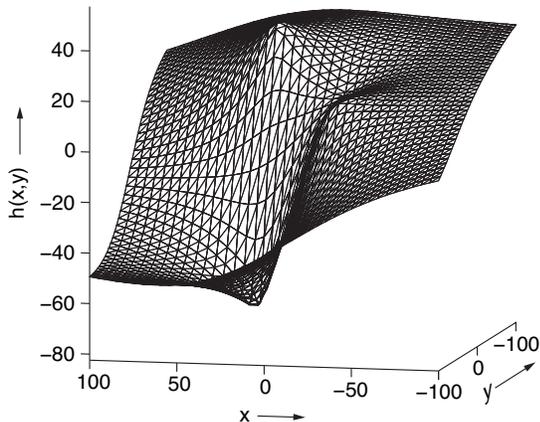}
%{c:/user/manus/figs/162-227.eps}
\caption{3D plot of a 4-node height profile with nodes at $(\pm
20,\pm 20)$ and charges $2.0$, $-1.5$, $-1.0$, and $0.5$ (in units
of $2\nu/\lambda$)} \label{fig2}
\end{figure}

The construction of the network is implemented in terms of the
slope field ${\bf u}$. Assigning a dilute network of static modes
at positions ${\bf r}_i^0, i=1\cdots$ with charges $k_i$ the total
slope field is ${\bf u}({\bf r})=(2\nu/\lambda)\sum_i\nabla
w_i(|{\bf r}-{\bf r}_i^0|)/w_i(|{\bf r}-{\bf r}_i^0|)$, $\sum_i
k_i=0$. In the vicinity of the mode position ${\bf r}_l^0$ the
slope field is shifted by ${\bf u}^0_l= (2\nu/\lambda)\sum_{i\neq
l}\nabla w_i(|{\bf r}_l^0-{\bf r}_i^0|)/w_i(|{\bf r}_l^0-{\bf
r}_i^0|)$ and the $l$-th mode is assigned the velocity ${\bf
v}_l=-\lambda{\bf u}^0_l$. Using the asymptotic expressions for
the modes and introducing a core radius $\epsilon$ or order
$k^{-1}$ the {\it self-consistent} dilute dynamical network
constituting the growth of an interface together with the
associated action is given by
\begin{eqnarray}
&&h({\bf r},t)=\frac{2\nu}{\lambda}\sum_i k_i \sqrt{({\bf r}-{\bf
r}_i(t))^2+\epsilon},
\\
&&{\bf r}_i(t)=\int_0^t{\bf v}_i(t')dt'+{\bf r}_i^0,~~\sum_ik_i=0,
\\
&&{\bf v}_i~(t)=-2\nu\sum_{l\neq i}k_l\frac{{\bf r}_i(t)-{\bf
r}_l(t)} {|{\bf r}_i(t)-{\bf r}_l(t)|},
\\
&&S=\sum_{i,k_i<0} S_i,~~S_i=\frac{k_0^2}{2}T\int d^dx~w_i^4.
\end{eqnarray}
In Fig.~\ref{fig2} we depict a 3D snapshot of a 4-node growth
morphology in $d=2$.

In the linear case for $k_0=0$ the equations of motion
(\ref{eq11}) and (\ref{eq22}) admit extended diffusive modes with
dispersion $\omega=\nu k^2$, corresponding to the EW universality
class, as discussed in $d=1$ case in Refs.
\cite{Fogedby99a,Fogedby03b}. The linear modes do not contribute
to the growth and are superimposed on the dynamical network in the
KPZ case.

On the basis of mode coupling and DP theory
\cite{Moore95,Laessig97,Colaiori01a} it has been argued that the
upper critical dimension for the KPZ equation is $d_c=4$. In the
present context $d_c$ is associated with the absence of growth
modes above $d=4$ as indicated by the numerics of the NLSE. Here I
present a more rigorous argument for the existence of an upper
critical dimension using Derrick's theorem
\cite{Derrick64,Rasmussen86} based on constrained minimization.

The NLSE (\ref{nlse}) can be derived from the variation of the
free energy $F=K+(1/2)k^2N-k_0^2I$, where $K=(1/2)\int
d^dx~(\nabla w)^2$ is the deformation energy, $N=\int d^dx~w^2$
the norm, and $I=\int d^dx~w^4$ the interaction, i.e., $\delta
F/\delta w=0$ yields $\nabla^2w=k^2w-k_0^2w^3$. Moreover,
multiplying the NLSE by $w$ and integrating over space yields the
first identity: $-2K=k^2N-k_0^2I$. Under the scale transformation
$w({\bf r})\rightarrow w(\mu{\bf r})$ one infers
$K\rightarrow\mu^{d-2}K$, $N\rightarrow\mu^dN$, and
$I\rightarrow\mu^dI$ and subject to constrained minimization
$dF/d\mu|_{\mu=1}=0$ the second identity:
$(d-2)K+(k^2/2)dN-(k^2_0/4)dI=0$. Eliminating $K$ from the
identities one infers $k^2N=(k^2_0/4)(4-d)I$. Since $N>0$ and
$I>0$ it follows that $d<4$ in order for a bound state to exist
with finite norm.

In the present Letter I have summarized a general nonperturbative
variationally-based weak noise approach to white noise driven
stochastic processes. The method has, moreover, been applied to
the KPZ equation in arbitrary dimensions yielding a Galilean
invariant self-consistent dynamical network of modes accounting
for the kinetic growth of the interface. In $d=1$ the results
agree with earlier findings, i.e, a network of matched domain
walls whose dispersion yields the dynamic exponent $z=3/2$; in
$d>1$, the detailed scaling properties remain to be worked out.
Finally, based on the dynamical network representation and
constrained minimization I have given an argument for $d=4$ as the
upper critical dimension for the KPZ equation.

Discussions with A. Svane and J. Krug  are gratefully
acknowledged. The present work has been supported by the Danish
Research Council.

%\bibliography{c:/user/manus/bib/articles,c:/user/manus/bib/books}
\end{document}